\documentclass[two column, prA]{revtex4}%
\usepackage{amsmath}
\usepackage{graphicx}
\usepackage{amsfonts}
\usepackage{amssymb}
\usepackage{color}%
\providecommand{\U}[1]{\protect\rule{.1in}{.1in}}
\ifx\pdfoutput\relax\let\pdfoutput=\undefined\fi
\newcount\msipdfoutput
\ifx\pdfoutput\undefined\else
\ifcase\pdfoutput\else
\ifx\paperwidth\undefined\else
\ifdim\paperheight=0pt\relax\else\pdfpageheight\paperheight\fi
\ifdim\paperwidth=0pt\relax\else\pdfpagewidth\paperwidth\fi
\fi\fi\fi
\begin{document}
\title[ ]{Conserved photon current }
\author{Margaret Hawton}
\affiliation{Lakehead University, Thunder Bay, ON, Canada, P7B 5E1}
\email{mhawton@lakeheadu.ca}

\begin{abstract}
A conserved photon current is derived from the commutation relations satisfied
by the electromagnetic four-potential and field tensor operators. The density
is found to be a sum over positive and negative frequency terms, both of which
contribute a positive number density and propagate in a common direction.
Discrete positive and negative frequency excitations are both identified as
photons. Photon number, equal to the spatial integral of photon density, is
conserved in the absence of sources and sinks.

\end{abstract}
\maketitle

\section{Introduction}

In quantum field theory (QFT) and quantum optics (QO) photons are indivisible
excitations of the electromagnetic (EM) field with integral helicity
$\lambda=\pm1$, making them bosons described by second quantized fields
satisfying commutation relations. Since there is no exclusion principle for
bosons, an EM state is an arbitrary linear combination of $n$-photon states
where $n$ is any positive integer or $0$; it is a whole number. We will refer
to both positive and negative frequency EM excitations as photons.

Physical one-photon pulses coupled to transmission lines and optical circuits
are now routinely prepared in the laboratory
\cite{LounisOrritSinglePhotonSources}. A one-photon pulse with bandwidth equal
to an appreciable fraction of its center frequency has definite photon number
but not definite energy. Photon density is fundamental and the scalar product
that normalizes a one-photon state should be conserved in the absence of
sources and sinks. 

The model described here can be applied to the essential components of an
optical quantum computor. In 2000 Knill, Laflamme and Milburn proved that it
is possible to create a universal quantum computer solely with single photon
sources, optical gates consisting of beam splitters and phase shifters and
photodetectors (KLM protocol) \cite{KLM}. Photons are ideal quantum devices
\cite{RM} and a photonic integrated circuit (PIC) can be implemented using
established foundary based technology \cite{Manufacturable}. These devices can
incorporate photon sources, low loss dielectric transmission lines, optical
gates and photon counting detectors \cite{Multiphoton,UltraLowLoss}. The
transmission lines and optical gates will be referred to here as the optical circuit.

The source, optical circuit and detector should be, at least approximately,
confined to separable finite regions of space. This degree of localization
cannot be described by a positive frequency field alone since any positive
frequency function initially localized in a finite region spreads
instantaneously throughout space \cite{Hegerfeldt} and there are no local
annihilation or creation operators \cite{RS}.\ The early work on approximate
photon localization is reviewed by Mandel and Wolf \cite{MandlWolf} and use of
Hermitian operators to describe electromagnetic excitation in QFT is discussed
in \cite{GellMann}. 

Motivated by these no-go theorems, previous work on quantization of positive
and negative frquency photon states \cite{HawtonDebierre} and the need to
model localized devices such as beam splitters, researchers at the university
of Leeds quantized both positive and negative frequency solutions of Maxwell's
equation in position space to give real localizable photon pulses
\cite{Leeds}. Consistent with these requirements and previous work, the
conserved photon current described here is also a real localizable sum of
positive and negative frequency fields.

The conserved electric current is well known and serves as a model for the
quantitative description of a conserved photon current. Since $\mathcal{F}%
^{\mu\nu}=\partial^{\mu}A^{\nu}-\partial^{\nu}A^{\mu}$ is a tensor its
contraction with $A_{\mu}$,
\begin{equation}
A_{\mu}\mathcal{F}^{\mu\nu}=\left(  \frac{1}{c}\mathbf{A}\cdot\mathbf{E}%
,\;\mathbf{A}\times\mathbf{B}+\frac{1}{c^{2}}\mathbf{E}\phi\right)
,\label{AF}%
\end{equation}
is a four-vector suggesting a candidate for the conserved photon current. In
the next section we will calculate the commutator $\frac{i\varepsilon_{0}%
}{2\hbar}\left[  \widehat{A}_{\mu},\widehat{\mathcal{F}}^{\mu\nu}\right]  $
and show that it describes a real localizable four-current operator satisfying
a continuity equation with material source and sink current $\widehat{J}_{e}$.
In the final Section its relationship to one-photon quantum mechanics and
experiment will be discussed.

\section{Conserved photon current}

We first define the notation: SI units will be used throughout. The
contravariant space-time, wave vector and momentum four-vectors are $x=x^{\mu
}=\left(  ct,\mathbf{x}\right)  ,$ $k=\left(  \omega_{k}/c,\mathbf{k}\right)
$ and $p=\hbar k$ where $kx=\omega_{k}t-\mathbf{k\cdot x}$ is invariant, the
four-gradient is $\partial=\left(  \partial_{ct},-\mathbf{\nabla}\right)  $,
$\square\equiv\partial_{\mu}\partial^{\mu}=\partial_{ct}^{2}-\nabla^{2}$, the
four-potential is $A\left(  t,\mathbf{x}\right)  =A^{\mu}=\left(  \frac{\phi
}{c},\mathbf{A}\right)  ,$ and a four-current is $J^{\mu}=\left(  \rho
c,\mathbf{J}\right)  $. The covariant four-vector corresponding to $U^{\mu
}=\left(  U_{0},\mathbf{U}\right)  $ is $U_{\mu}=g_{\mu\nu}U^{\nu}=\left(
U_{0},-\mathbf{U}\right)  $ where $g_{\mu\nu}=g^{\mu\nu}$ is a $4\times4$
diagonal matrix with diagonal $\left(  1,-1,-1,-1\right)  $ and $U_{\mu}%
U^{\mu}=U^{\mu}U_{\mu}$ is an invariant. The mutually orthogonal unit vectors
$e^{\mu}$ are defined such that $e_{0}=n^{\mu}=\left(  1,0,0,0\right)  $ is
time-like, $\mathbf{e}_{\mathbf{k}}=\mathbf{k}/\left\vert \mathbf{k}%
\right\vert $ is longitudinal, and the definite helicity transverse Lorentz
invariant unit vectors are $\mathbf{e}_{\lambda}\left(  \mathbf{k}\right)  $
with $\lambda=\pm1$.

The classical electric and magnetic fields%
\begin{equation}
\mathbf{E}=-\partial_{t}\mathbf{A}-\nabla\phi,\;\mathbf{B}=\nabla
\times\mathbf{A}\label{EandBclassical}%
\end{equation}
can be written in covariant form as the Faraday tensor%
\begin{equation}
\mathcal{F}^{\mu\nu}=\partial^{\mu}A^{\nu}-\partial^{\nu}A^{\mu}=\frac{1}%
{c}\left(
\begin{array}
[c]{cccc}%
0 & -E_{x} & -E_{y} & -E_{z}\\
E_{x} & 0 & -cB_{z} & cB_{y}\\
E_{y} & cB_{z} & 0 & -cB_{x}\\
E_{z} & -cB_{y} & cB_{x} & 0
\end{array}
\right)  \label{F}%
\end{equation}
that is antisymmetric under exchange of indices that is%
\begin{equation}
\mathcal{F}^{\mu\nu}=-\mathcal{F}^{\nu\mu}.\label{antisym}%
\end{equation}
The Standard Lagrangian density is
\begin{equation}
\mathcal{L}=-\frac{1}{4}\varepsilon_{0}c^{2}\mathcal{F}_{\mu\nu}%
\mathcal{F}^{\mu\nu}-J_{e}^{\mu}A_{\mu}\label{L}%
\end{equation}
where $-\frac{1}{4}\varepsilon_{0}c^{2}\mathcal{F}_{\mu\nu}\mathcal{F}^{\mu
\nu}=\frac{1}{2}\varepsilon_{0}\left(  \mathbf{E}\cdot\mathbf{E}%
-c^{2}\mathbf{B}\cdot\mathbf{B}\right)  $ and $J_{e}^{\mu}$ is the conserved
electric four-current. $\text{The Lagrangian is }L=\int d\mathbf{x}%
\mathcal{L}\left(  t,\mathbf{x}\right)  $. The analog of the position
coordinate in this Lagrangian is $A^{\mu}\left(  x\right)  $ so the
four-momentum conjugate to $A^{\nu}$ is $\varepsilon_{0}c\mathcal{F}^{0\nu}%
$\textbf{.} After second quantization the form of the classical equations of
motion, $\varepsilon_{0}c^{2}\partial_{\mu}\mathcal{F}^{\mu\nu}=J_{e}^{\mu},$
is preserved so the photon field operators satisfy%
\begin{equation}
\varepsilon_{0}c^{2}\partial_{\mu}\widehat{\mathcal{F}}^{\mu\nu}%
=\widehat{J}_{e}^{\mu}.\label{EqMotion}%
\end{equation}
The continuity equation for $\left[  \widehat{A}_{\mu},\widehat{\mathcal{F}%
}^{\mu\nu}\right]  $ is%
\begin{align}
\partial_{\nu}\left[  \widehat{A}_{\mu},\widehat{\mathcal{F}}^{\mu\nu}\right]
&  =\left[  \left(  \partial_{\nu}\widehat{A}_{\mu}\right)
,\widehat{\mathcal{F}}^{\mu\nu}\right]  +\left[  \widehat{A}_{\mu},\left(
\partial_{\nu}\widehat{\mathcal{F}}^{\mu\nu}\right)  \right]  ,\nonumber\\
\left[  \left(  \partial_{\nu}\widehat{A}_{\mu}\right)  ,\widehat{\mathcal{F}%
}^{\mu\nu}\right]   &  =\left(  \partial_{\nu}\widehat{A}_{\mu}\right)
\left(  \partial^{\mu}\widehat{A}^{\nu}\right)  -\left(  \partial_{\nu
}\widehat{A}_{\mu}\right)  \left(  \partial^{\nu}\widehat{A}^{\mu}\right)
\nonumber\\
&  -\left(  \partial^{\mu}\widehat{A}^{\nu}\right)  \left(  \partial_{\nu
}\widehat{A}_{\mu}\right)  +\left(  \partial^{\nu}\widehat{A}^{\mu}\right)
\left(  \partial_{\nu}\widehat{A}_{\mu}\right)  \nonumber\\
&  =0,\nonumber\\
\partial_{\nu}\left[  \widehat{A}_{\mu},\widehat{\mathcal{F}}^{\mu\nu}\right]
&  =\frac{1}{\varepsilon_{0}c^{2}}\left[  \widehat{A}_{\mu},\widehat{J}%
_{e}^{\mu}\right]  \label{continuity}%
\end{align}
where covariant and contravariant indices were exchanged to prove the second
equation and (\ref{EqMotion}) was used to give the final result,
(\ref{continuity}). The equations of motion (\ref{EqMotion}) are second
quantized versions of the MEs $\mathbf{\nabla}\cdot\mathbf{E}=\rho
_{e}/\varepsilon_{0},~\partial_{t}\mathbf{E}-c^{2}\mathbf{\nabla}%
\times\mathbf{B=}-\varepsilon_{0}^{-1}\mathbf{J}_{e}$.

The four-potential $A_{\mu}$ in (\ref{continuity}) is gauge dependent. Eq.
(\ref{continuity}) describes all second quantized propagating fields. In the
Coulomb gauge only transverse modes are second quantized, while in the Lorenz
gauge transverse, longitudinal and scalar modes propagate. The Lorenz gauge
requires addition of the gauge term $-\frac{1}{2}\varepsilon_{0}c^{2}%
\Lambda^{2}$ with $\Lambda=c^{-2}\partial_{t}\phi+\mathbf{\nabla}%
\cdot\mathbf{A}$ to (\ref{L}) to give momentum conjugate to $\phi$ equal to
$\varepsilon_{0}\partial_{t}\phi/c$. The scalar product is indefinite and the
equation of motion (\ref{EqMotion}) is modified \cite{CT}. It is sufficient
for the applications to be discussed here to consider only transverse electric
fields $\mathbf{E=E}_{\perp}$ that propagate in free space and remain
transverse when transmitted into a dielectric. In this case $\mathbf{E}%
_{\parallel}=-\partial_{t}\mathbf{A}_{\parallel}-\nabla\phi=0$ gives
$\mathbf{A}_{\parallel}\times\mathbf{B=-E}_{\perp}\phi/c^{2}$ so%
\begin{align}
A_{\mu}\mathcal{F}_{\perp}^{\mu\nu}  & =\left(  \frac{1}{c}\mathbf{A}%
\cdot\mathbf{E}_{\perp},\;\mathbf{A}\times\mathbf{B}+\mathbf{E}_{\perp}%
\phi\right)  \nonumber\\
& =\left(  \frac{1}{c}\mathbf{A}_{\perp}\cdot\mathbf{E}_{\perp},\mathbf{A}%
_{\perp}\times\mathbf{B}\right)  \label{Gauge}%
\end{align}
is gauge independent and remains gauge independent when second quantized. 

Starting with the positive frequency annihilation operator
$\widehat{\mathbf{A}}_{\lambda}^{+}\left(  x\right)  $ the remaining EM
creation and annihilation operators can be calculated. The positive frequency
operators are second quantized versions of the classical analytic signal.
These operators provide a convenient mathematical description of propagation
in the optical circuit with phase shifters $e^{i\phi}$ and beam splitters with
complex reflectivity and transmisivity $r$ and $t$ that satisfy $\left\vert
r\right\vert ^{2}+\left\vert t\right\vert ^{2}=1$ and hence conserve photon
number. The remaining field operators are
\begin{align}
\ \widehat{\mathbf{A}}^{-}\left(  x\right)   &  =\widehat{\mathbf{A}%
}^{+\dagger}\left(  x\right)  ,\label{A-}\\
\widehat{\mathbf{A}}\left(  x\right)   &  =\sum_{\lambda=\pm1}\left(
\widehat{\mathbf{A}}_{\lambda}^{+}\left(  x\right)  +\widehat{\mathbf{A}%
}_{\lambda}^{-}\left(  x\right)  \right)  ,\label{A}\\
\widehat{\mathbf{E}}_{\perp}\left(  x\right)   &  =-\partial_{t}%
\widehat{\mathbf{A}}_{\perp}\left(  x\right)  ,\ \widehat{\mathbf{B}}\left(
x\right)  =\mathbf{\nabla\times}\widehat{\mathbf{A}}\left(  x\right)
.\label{EandB}%
\end{align}

Since annihilation and creation operators commute amongst themselves
\begin{equation}
\left[  \widehat{A}_{\mu},\widehat{\mathcal{F}}_{\perp}^{\mu\nu}\right]
=\left[  \widehat{A}_{\mu}^{+},\widehat{\mathcal{F}}_{\perp}^{\mu\nu-}\right]
+\left[  \widehat{A}_{\mu}^{-},\widehat{\mathcal{F}}_{\perp}^{\mu\nu+}\right]
.\label{Comm+-}%
\end{equation}
The photon current density operator array%
\begin{align}
\widehat{J}_{p12}^{\lambda\lambda^{\prime}}\left(  x,x^{\prime}\right)   &
=\frac{-i\varepsilon_{0}}{2\hbar}\left(  \widehat{\mathbf{A}}_{2}^{\lambda
+}\left(  x\right)  \cdot\widehat{\mathbf{E}}_{1}^{\lambda^{\prime}-}\left(
x^{\prime}\right)  \right.  -\nonumber\\
&  \left.  \left.  -\widehat{\mathbf{E}}_{1}^{\lambda^{\prime}+}\left(
x^{\prime}\right)  \cdot\widehat{\mathbf{A}}_{2}^{\lambda-}\left(  x\right)
,\ \right.  \widehat{\mathbf{A}}_{2}^{\lambda+}\left(  x\right)  \times
c\widehat{\mathbf{B}}_{1}^{\lambda^{\prime}-}\left(  x^{\prime}\right)
\right.  \nonumber\\
&  \left.  -c\widehat{\mathbf{B}}_{1}^{\lambda^{\prime}+}\left(  x\right)
\times\widehat{\mathbf{A}}_{2}^{\lambda+}\left(  x^{\prime}\right)  \right)
\label{Jarray}%
\end{align}
generalizes (\ref{continuity}) to describe creation and annihilation of
photons at different space-time points for different, possibly orthogonal,
states. The generalization to modes $1$ and $2$ is only included for
convenience in defining the scalar product. The current density operator
(\ref{Jarray}) describes the addition of one photon to any Fock state. 

The source free MEs are space-time reversal invariant but emission and
detection of a photon in the laboratory is not. It will be assumed that
$t>t^{\prime}$ in the laboratory frame so creation of a photon at $x$ with
annihilation at $x^{\prime}$ will be interpreted as propagation of an
antiphoton from $x^{\prime}$ to $x$. The $\mathbf{A}^{+}\left(  x\right)
\cdot\mathbf{E}^{-}\left(  x^{\prime}\right)  $ term describes propagation of
a photon from space-time point $x^{\prime}$ to $x$, while the $\mathbf{E}%
^{+}\left(  x^{\prime}\right)  \cdot\mathbf{A}^{-}\left(  x\right)  $ term is
equivalent to an antiphoton propagating from $x^{\prime}$ to $x$. Both photons
and antiphotons that propagate from $x^{\prime}$ to $x$ in the laboratory
frame will be counted as photons. Since the minus sign in the commutation
relation is cancelled by the sign of the space-time derivatives of
$\mathbf{A}$ in $\mathbf{E}$ and $\mathbf{B}$ in (\ref{Jarray}), density is
positive and propagation is in a common direction for both its terms.

The generalized photon number density operator is%
\begin{align}
\widehat{\rho}_{p12}^{\lambda\lambda^{\prime}}\left(  x,x^{\prime}\right)    &
=\frac{i\varepsilon_{0}}{2\hbar}\left(  \widehat{\mathbf{A}}_{2}^{\lambda
+}\left(  x\right)  \cdot\widehat{\mathbf{E}}_{1}^{\lambda^{\prime}-}\left(
x^{\prime}\right)  \right.  \nonumber\\
& \left.  -\widehat{\mathbf{E}}_{1}^{\lambda^{\prime}+}\left(  x^{\prime
}\right)  \cdot\widehat{\mathbf{A}}_{2}^{\lambda-}\left(  x\right)  \right)
\label{number_density_op}%
\end{align}
Defining the scalar product in the normalized zero-photon state as
$\left\langle 0|0\right\rangle =1$, the one-photon scalar product on the $t$
hyperplane is%
\begin{align}
\rho_{p12}^{\lambda\lambda^{\prime}} &  =\left\langle 0\left\vert
\widehat{\rho}_{p12}^{\lambda\lambda^{\prime}}\left(  x,x\right)  \right\vert
0\right\rangle \label{scalarprod}\\
&  =\frac{i\varepsilon_{0}}{2\hbar}\int d\mathbf{x}\left(  \mathbf{A}%
_{2}^{\lambda+}\left(  x\right)  \cdot\mathbf{E}_{1}^{\lambda^{\prime}%
-}\left(  x\right)  -\widehat{\mathbf{E}}_{1}^{\lambda^{\prime}+}\left(
x\right)  \cdot\mathbf{A}_{2}^{\lambda-}\left(  x\right)  \right)  \nonumber
\end{align}
where $\mathbf{A}_{2}^{\lambda}\left(  x\right)  $ and $\mathbf{E}%
_{1}^{\lambda^{\prime}}\left(  x\right)  $ are one-photon EM fields with
helicities $\lambda$ and $\lambda^{\prime}$ respectively. Since $i\times$ the
integrand in (\ref{scalarprod}) is real its localizability is not limited by
the Hegerfeldt theorem. 

In QFT it is convertional to define a plane wave basis localized in
$\mathbf{k}$-space and a space-time basis that is localized in position space.
Here we follow the derivation of Fock space in \cite{SZ} by starting with the
periodic boundary conditions $k_{i}L=2\pi l_{i}$ for integral $l_{i}$ with
$i=x,y,z$ in volume $V=L^{3}$ and then taking the $V\rightarrow\infty$ limit.
The commutation relations will be written as
\begin{equation}
\left[  \widehat{a}_{\lambda\mathbf{k}},\widehat{a}_{\lambda^{\prime
}\mathbf{k}^{\prime}}^{\dag}\right]  =\delta_{\lambda\lambda^{\prime}}%
\delta_{\mathbf{k,k}^{\prime}}.\label{k_commutation}%
\end{equation}
Defining the $n$-photon annihilation operator
\begin{equation}
\widehat{a}_{\lambda\mathbf{k}n}\equiv\frac{\left(  \widehat{a}_{\lambda
\mathbf{k}}\right)  ^{n}}{\sqrt{n!}}\label{n_photon_annihilation}%
\end{equation}
it can be verified using the commutation relations (\ref{k_commutation}) that
\begin{equation}
\left\vert a_{\lambda\mathbf{k}n}\right\rangle =\widehat{a}_{\lambda
\mathbf{k}n}\left\vert 0\right\rangle \label{psi_n}%
\end{equation}
are the normalized $n$-photon Fock states. The number of states per unit
volume for a photon with definite helicity is
\begin{equation}
\underset{V\rightarrow\infty}{lim}\frac{\Delta\mathbf{n}}{V}=\frac
{d\mathbf{k}}{\left(  2\pi\right)  ^{3}}\label{density}%
\end{equation}
so $V^{-1}\sum_{\mathbf{k}}\rightarrow\left(  2\pi\right)  ^{-3}\int%
_{t}d\mathbf{k}$ and the scalar product%

\begin{equation}
\left[  \widehat{a}_{\lambda}\left(  \mathbf{k}\right)  ,\widehat{a}%
_{\lambda^{\prime}}^{\dagger}\left(  \mathbf{k}^{\prime}\right)  \right]
=\delta_{\lambda\lambda\prime}\delta\left(  \mathbf{k-k}^{\prime}\right)
\label{k_commutator}%
\end{equation}
defines a basis of orthonormal states. Since $\int\frac{d\mathbf{k}}%
{\omega_{k}}$ is an invariant, $\omega_{k}\delta\left(  \mathbf{k-k}^{\prime
}\right)  $ and $\omega_{k}^{1/2}\widehat{a}_{\lambda}\left(  \mathbf{k}%
\right)  $ are invariant. The operator
\begin{equation}
\widehat{n}_{\lambda}\left(  \mathbf{k}\right)  =\widehat{a}_{\lambda}^{\dag
}\left(  \mathbf{k}\right)  \widehat{a}_{\lambda}\left(  \mathbf{k}\right)
\label{count}%
\end{equation}
counts photons with wave vector $\mathbf{k}$ and helicity $\lambda$. This is
the Schr\H{o}dinger picture. In the Heisenberg picture the positive frequency
plane wave annihilation operator $\widehat{a}_{\lambda}\left(  \mathbf{k}%
,t\right)  =\widehat{a}_{\lambda}\left(  \mathbf{k}\right)  e^{-i\omega_{k}t}$
satisfies $i\partial_{t}\widehat{a}_{\lambda}\left(  \mathbf{k},t\right)
=\omega_{k}\widehat{a}_{\lambda}\left(  \mathbf{k},t\right)  $.

In the plane wave basis a general positive frequency vector potential operator
is%
\begin{equation}
\widehat{\mathbf{A}}_{\lambda}^{+}\left(  x\right)  =i\sqrt{\frac{\hbar
}{2\varepsilon_{0}}}\int\frac{d\mathbf{k}}{\left(  2\pi\right)  ^{3/2}%
\omega_{k}^{1/2}}c_{\lambda}\left(  \mathbf{k}\right)  \widehat{a}_{\lambda
}\left(  \mathbf{k}\right)  \mathbf{e}_{\lambda}\left(  \mathbf{k}\right)
e^{-ikx}.\label{A+}%
\end{equation}
The photon number (\ref{number}) for a state with helicity $\lambda$ is%
\begin{align}
n_{p}^{\lambda}  & =\frac{i\varepsilon_{0}}{2\hbar}\int d\mathbf{x}\left(
\mathbf{A}_{2}^{\lambda+}\left(  x\right)  \cdot\mathbf{E}_{1}^{\lambda
-}\left(  x\right)  -\widehat{\mathbf{E}}_{1}^{\lambda+}\left(  x\right)
\cdot\mathbf{A}_{2}^{\lambda-}\left(  x\right)  \right)  \label{number}\\
& =\int\frac{d\mathbf{k}}{\left(  2\pi\right)  ^{3}}\left\vert c_{\lambda
}\left(  \mathbf{k}\right)  \right\vert ^{2}\label{n_kspace}%
\end{align}
with mode $2$ equal to mode $1$ omitted from the notation now that the scalar
product (\ref{scalarprod}) has been defined. For a one-photon state
$n_{p}^{\lambda}=1$. 

If $c_{\lambda}\left(  \mathbf{k}\right)  =e^{ikx^{\prime}}$ expressions
(\ref{EandB}) and (\ref{k_commutation}) to (\ref{k_commutator}) substituted
into (\ref{Jarray}) give%
\begin{equation}
J_{p}^{\lambda\lambda^{\prime}}\left(  x,x^{\prime}\right)  =\frac{1}{2}%
\int\frac{d\mathbf{k}}{\left(  2\pi\right)  ^{3}}\delta_{\lambda
\lambda^{\prime}}\left(  1,\mathbf{e}_{\lambda}\left(  \mathbf{k}\right)
\right)  e^{-ik\left(  x-x^{\prime}\right)  }+c.c.\label{J12}%
\end{equation}
where $c.c.$ is the complex conjugate. Its zero component is the photon
density
\begin{equation}
\rho_{p}^{\lambda\lambda^{\prime}}\left(  x,x^{\prime}\right)  =\frac{1}%
{2}\int\frac{d\mathbf{k}}{\left(  2\pi\right)  ^{3}}\delta_{\lambda
\lambda^{\prime}}e^{ik\left(  x-x^{\prime}\right)  }+c.c..\label{QEDcomm}%
\end{equation}
On the $t=t^{\prime}$ hyperplane
\begin{equation}
\rho_{p}^{\lambda\lambda^{\prime}}\left(  x,x^{\prime}\right)  =\delta
_{\lambda\lambda^{\prime}}\delta\left(  \mathbf{x}^{\prime}-\mathbf{x}\right)
\label{x_basis}%
\end{equation}
so (\ref{A+}) describes an orthonormal basis of localized states. The
operators $\widehat{\mathbf{A}}\left(  t,\mathbf{x}\right)  $ and
$\widehat{\mathbf{E}}\left(  t,\mathbf{x}^{\prime}\right)  $ commute for
spacelike separated points $\mathbf{x}^{\prime}\mathbf{\neq x}$, so a
measurement at $\mathbf{x}^{\prime}$ does not change the outcome at
$\mathbf{x}$. This enforces causality in QED. Using (\ref{density}), the zero
component of (\ref{J12}) can be written as%
\begin{equation}
\rho_{p}^{\lambda\lambda^{\prime}}\left(  x,x^{\prime}\right)  =\frac{1}%
{2}\sum_{\mathbf{k}}\delta_{\lambda\lambda^{\prime}}\frac{e^{ik\left(
x-x^{\prime}\right)  }}{V}+c.c.,\label{rho_p}%
\end{equation}
verifying that at $x=x^{\prime}$ it is a density equal to a sum over plane
wave number densities.

With $\left\vert \Delta\mathbf{x}\right\vert \equiv\left\vert \mathbf{x-x}%
^{\prime}\right\vert $ and $\Delta t\equiv t-t^{\prime}$ (\ref{QEDcomm}) gives%
\begin{equation}
\rho_{p}^{\lambda\lambda^{\prime}}\left(  x,x^{\prime}\right)  =\rho
_{p}^{\lambda\lambda^{\prime}+}\left(  x,x^{\prime}\right)  +\rho_{p}%
^{\lambda\lambda^{\prime}+\ast}\left(  x,x^{\prime}\right)  \label{rho_p_real}%
\end{equation}
where the positive frequency part of $\rho_{p}^{\lambda\lambda^{\prime}}$ is
\begin{align}
\rho_{p}^{\lambda\lambda^{\prime}+}\left(  x,x^{\prime}\right)   &  =\int%
\frac{d\mathbf{k}}{2\left(  2\pi\right)  ^{3}}e^{-ik\left(  x-x^{\prime
}\right)  }\delta_{\lambda\lambda^{\prime}}\nonumber\\
&  =\frac{-1}{8\pi^{2}r}\frac{\partial}{\partial r}\left(  \pi\delta\left(
\left\vert \Delta\mathbf{x}\right\vert -c\Delta t\right)  \right.  \nonumber\\
&  \left.  +iP\left(  \frac{1}{\left\vert \Delta\mathbf{x}\right\vert -c\Delta
t}\right)  \right)  \delta_{\lambda\lambda^{\prime}}\label{MaxwellRS}%
\end{align}
and $P$ is the principal value. The positive and negative frequency
contributions to photon number density are separately nonlocal, but their sum
is real and localized on a spherical shell. The three-dimensional case can be
used to model emission from a localized atom or quantum dot initially in an
excited state.

For a one dimensional photon pulse with spatially uniform area $A$
\ propagating in the $+k_{x}$ direction the positive frequency part of the
photon density is%
\begin{align}
\rho_{p}^{\lambda\lambda^{\prime}+}\left(  x,x^{\prime}\right)   &  =\int%
_{0}^{\infty}\frac{dk_{x}}{2\pi A}e^{ik_{x}\left(  x-x^{\prime}\right)
}\delta_{\lambda\lambda^{\prime}}\nonumber\\
&  =\frac{1}{2A}\left(  \delta\left(  \Delta x-c\Delta t\right)  \right.
\nonumber\\
&  -\left.  \frac{i}{\pi}P\left(  \frac{1}{\Delta x-c\Delta t}\right)
\right)  \delta_{\lambda\lambda^{\prime}}\label{rho1D}%
\end{align}
where $\Delta t\equiv t-t^{\prime}$, $\Delta x\equiv x-x^{\prime}$ and
\begin{equation}
\rho_{p}^{\lambda\lambda^{\prime}+}+\rho_{p}^{\lambda\lambda^{\prime}+\ast
}=\frac{1}{A}\delta\left(  \Delta x-c\Delta t\right)  \delta_{\lambda
\lambda^{\prime}}\label{localized_ppbar}%
\end{equation}
is $\delta$-localized and propagates at speed $c$. This localized basis can be
integrated over to describe localization in a finite region. An example of
instantaneous localization in a square well is given in \cite{Prigogine}.

In classical EM the macroscopic description of transmission and reflection at
a dielectric interface is know to work for visible and infrared light. This is
based on averaging over domains with dimensions of order $10^{-8}m$ that
include many molecules\ \cite{Jackson}. This classical macroscopic model will
be second quantized to model photon propagation in a dielectric such as a
transmission line. The photon source and photon counting detectors are
separate devices whose details are not considered here, it is only assumed
that the sources emit single photons and the detectors are photon counting
devices. 

Only whole numbers of EM excitations exist - there are no fractional photons.
This was verified experimentally in \cite{BeamSplitter} where a detector was
placed in two paths and, within experimental error, no coincident photon
detection events were observed.\ The photon number density in free space is
$\varepsilon_{0}\mathbf{E}\cdot\mathbf{A}$. When a one-photon pulse passes
from free space into a dielectric material or optical circuit photon number
must remain $n=1$ until the photon is absorbed in a lossy material or counted
in an optical detector and reduced to the $n=0$ state. To preserve the
normalization $\int d\mathbf{x}\varepsilon_{0}\mathbf{E}\cdot\mathbf{A}=1$,
photon density must be $\varepsilon_{0}\mathbf{E}\cdot\mathbf{A}$ in the
dielectric. 

In a polarizable dielectric the transverse positive frequency second quantized
operators satisfy the MEs%
\begin{equation}
\partial_{t}\left(  \varepsilon_{0}\widehat{\mathbf{E}}^{+}%
+\widehat{\mathbf{P}}^{+}\right)  -\bigtriangledown\times\widehat{\mathbf{H}%
}^{+}\mathbf{=}-\widehat{\mathbf{J}}^{+s}\label{MEs}%
\end{equation}
where $\widehat{\mathbf{H}}^{+}\mathbf{=\mu}_{0}^{-1}\widehat{\mathbf{B}}%
^{+}=\varepsilon_{0}c^{2}\widehat{\mathbf{B}}^{+}$ and, for simplicity, the
material has been assumed to be nonmagnetic. The current $\widehat{J}_{e}$ in
(\ref{continuity}) is driven by the electric field operator so it is also
operator-valued. A single-photon pulse transmitted into a transparent medium
must remain normalized so its number density remains $\varepsilon
_{0}\mathbf{E}\cdot\mathbf{A}$ as in free space. 

When a light pulse with momentum $\mathbf{p}_{em}$ propagating in free space
encounters a planar interface of a diectric with index of refraction $n$,
$r=\frac{n-1}{n+1}$ and $t=\frac{2n}{n+1}$ are determined by the Fresnel
equations. Total momentum is conserved so if this pulse is reflected off an
ideal mirror with reflectivity $r=1$ the mirror will gain momentum
$2\mathbf{p}_{em}$ and if it is absorbed the dielectric slab will gain
momentum $\mathbf{p}_{em}$ \cite{Resolution}. A one-photon pulse incident on
this ideal mirror it will remain a one-photon pulse when reflected and will be
reduced to the $n=0$ state if absorbed. 

A one-dimensional plane wave with helicity $\lambda$, $A_{\lambda k}^{+}%
=\exp\left[  i\omega_{k}\left(  x/c-t\right)  \right]  /2\pi\omega_{k}^{1/2}$,
incident from free space on a weakly absorbing dielectric medium with index of
refraction $n$ will be transmitted with probability amplitude $t$. The index
of refraction is in general complex with real and imaginary parts $n^{\prime
}\left(  \omega\right)  $ and $n^{\prime\prime}\left(  \omega\right)  ,$ so%
\begin{equation}
n\left(  \omega\right)  =\sqrt{1+\chi\left(  \omega\right)  }=n^{\prime
}\left(  \omega\right)  +in^{\prime\prime}\left(  \omega\right)  \label{n}%
\end{equation}
and, in the dielectric,%
\begin{equation}
A_{\lambda k}^{+}\left(  x,t\right)  =\frac{t}{2\pi\omega_{k}^{1/2}}%
\exp\left(  -\omega_{k}x\frac{n^{\prime\prime}}{c}\right)  \exp\left(
i\omega_{k}\left(  x\frac{n^{\prime}}{c}-t\right)  \right)  .\label{At}%
\end{equation}
Since one-photon number density is $\varepsilon_{0}\mathbf{E}\cdot\mathbf{A}$
in the dielectric, its momentum is of the Abraham form, $\mathbf{p}%
_{A}=\mathbf{p}_{em}=\varepsilon_{o}\mathbf{E}\times\mathbf{B}$
\cite{Resolution,Jackson}. The Minkowski momentum $\mathbf{p}_{M}%
=\mathbf{p}_{em}+\chi\mathbf{p}_{em}$ includes the momentum due to
polarization of the dielectric medium. This acts as a drag force on the single
photon, reducing its speed from $c$ to $c/n^{\prime}$.  In a PIC photons
propagate in multiple transmission lines with essentially identical
characteristics and common dielectric susceptability with complex
reflectivity, $r$, and transmissivity, $t$, \ of the light pulses determined
by the angles of their intersections \cite{Multiphoton}. 

\section{ Discussion}

In Section II a conserved photon four-current was derived from the
potential-field commutation relations. The photon probability density was used
to define a scalar product that can form a basis for a first quantized theory
of single photons. Here this scalar product is derived from fundamental
principles according to which the minus from space-time differentiation is
cancelled out by minus from the commutation relations to give a positive
photon number density for both positive and negative frequency fields.
Previous definitions of the one-photon scalar product were limited to use of
nonlocalizable positive frequency fields \cite{BB} or, motivated by the
observation that experimental one-photon pulses can be modelled classically
\cite{Barnett}, use of the Mostafazadeh \cite{Mostafazadeh} sign of frequency
operator \cite{Validation}. In the latter case the scalar product
(\ref{scalar_product}), derived here from fundamental principles, was
constructed on an ad hoc basis. Propagation of highly-localized wave packets
that remain localized at all times is discussed in \cite{Raymer} and
\cite{LocalPhotons}.

In a quantum optical circuit single photons are injected into input modes of a
linear interferometer described by a unitary operator, $\widehat{U}$. Since
$\widehat{U}$ \ is unitary, photon number is conserved, consistent with the
conservation law derived in Section II. Propagation of photon pulses in
dielectric media and through a beam splitter is discussed in \cite{FJ}. Since
$\int d\mathbf{x\varepsilon}_{0}\mathbf{E}\cdot\mathbf{A}=1$ for a one-photon
state in free space, the free space form of the photon number density must be
preserved in a dielectric. The polarization induced in the medium by
$\mathbf{E}$ does not contribute to photon number density, instead it acts as
a drag force that reduces the propagation speed of the single-photon pulse.
Single photon momentum density is of the Abraham form, $\mathbf{p}%
_{em}=\varepsilon_{0}\mathbf{E\times A}$.

There are no photons in classical electromagnetic theory - the discrete
excitations that we call photons are created and annihilated by second
quantized operators. The localizable causally propagating photon number
density derived here determines the probabilty that the photon will be
counted. Photon density must necessarily be interpreted as a probability
density since photons are indivisible and can be counted only once as verified
experimentally in \cite{BeamSplitter}. We have identified a conserved photon
four-current operator that describes both positive and negative frequency EM
excitations. Their sum is positive, real, localizable in a finite region, and
propagates causally. Photon number is conserved in the absence of sources and
sinks. Single photon states are represented by normalized fields that collapse
to the zero photon state when the photon is counted. This is a purely quantum
effect described by second quantization - it has no counterpart in classical
EM. The one-photon density must be interpreted as a probability density.

\end{document}